# Optically Facet-Resolved Reaction Anisotropy in Two-Dimensional Transition Metal Dichalcogenides

Myeongin Song,[1] Haneul Kang,[1] Dan Rhodes,[2] Bumho Kim,[2] James Hone,[2] and Sunmin Ryu[1]*

[1]Department of Chemistry, Pohang University of Science and Technology (POSTECH), Pohang, Gyeongbuk 37673, Korea

[2]Department of Mechanical Engineering, Columbia University, New York, NY 10027, USA

* E-mail: sunryu@postech.ac.kr

## Abstract

Quantifying anisotropy in the chemical reactions of mesoscopic crystals has mostly resorted on the combination of electron microscopy and diffraction. In this work, we established crystal-facet-resolved kinetic measurements of oxidation reactions in 2D transition metal dichalcogenides (TMDs) using optical second-harmonic generation spectroscopy and scanning probe microscopy. We show the in-plane anisotropy of their bond-breaking reactions is governed by their structure and strongly material-dependent among four TMDs. The facet-resolved analysis directly revealed that the reactions proceed fastest (slowest) for chalcogen (metal)-terminated zigzag edges with armchair edges in the middle. The degree of the anisotropy inducing trigonal oxidation patterns was much higher in $MoS_2$ and $MoSe_2$ than $WS_2$ and $WSe_2$. Kinetic Wulff construction based on edge-specific reaction rates verified the material-dependent mesoscopic reaction patterns. We also show that the reactions are initiated at substrate-mediated defects located on the bottom and top surfaces of 2D TMDs.

**Introduction**

Surfaces are an essential part of functional materials because they accommodate various physical and chemical exchanges in the form of energy and charges at the interface with other chemical entities. Because of their high surface-atom fraction, this statement is most relevant to two-dimensional (2D) materials represented by graphene and transition metal dichalcogenides (TMDs). Facile charge exchange with adsorbed molecules across the surface enabled controlling charge density of 2D materials through chemical doping of electrons and holes.[1-6] Photogenerated excitonic energy is transferred from or to 2D materials via ultrafast dipolar quenching.[7-10] For many physicochemical processes, however, their atomically flat surfaces are much less active than their edges or point defects,[1, 11, 12] which constitute their 1D and 0D surfaces, respectively. These sub-2D surfaces also support novel phenomena and modification of material properties such as vacancy-induced mid-gap states,[13] edge-state-mediated magnetism,[14] edge-enhanced photoluminescence[15] and optical second-harmonic generation (SHG).[16]

For nanoscale control of edge structures, 2D crystals serve as a good testbed because of their sheet-like geometry that allows facile lithographic write-and-cut in addition to their high crystallinity and chemical stability. Besides the conventional lithographic methods for microfabrication, various chemical reactions have been explored for controlled patterning of graphene: thermal oxidation for circular patterns,[1] activated hydrogen plasma for zigzag-terminated hexagons,[17] and dynamic control between zigzag and armchair edges.[18] In this regard, the family of TMDs presents more possibilities of control as binary compounds. 2D $MoS_2$ was found to develop zigzag-terminated[19] triangular patterns upon thermal oxidation.[20, 21] Despite these studies, however, 2D TMDs await further investigations for chemical manipulation of edges and defects, and understanding the underlying principles pertaining to 2D material systems. Most of all, the material dependence of given reactions has yet to be comprehended for custom-tailored patterning. Moreover, one needs to understand the reaction kinetics at edges that may vary as a function of their crystallographic orientation. Further insight is also required into the roles of atomic-level defects and interfaces with substrates in the initiation and propagation of the chemical reactions.

In this work, we report material-dependent anisotropy in the thermal oxidation of four 2D TMDs by establishing edge-resolved kinetic measurements based on optical second-harmonic generation (SHG) spectroscopy. W dichalcogenides exhibited rounded triangles of W oxides unlike Mo dichalcogenides showing equilateral triangular etch pits. The material-dependent anisotropy and branching between the oxide formation and etching are governed by the energy difference between competing edges and the thermal stability of oxides, respectively. We also show that the reactions are initiated at substrate-mediated defects that are located on the bottom and top surfaces of 2D TMDs.

**Methods**

**Preparation and Treatments of Samples.** 2D samples of four TMDs ($MoS_2$, $MoSe_2$, $WS_2$, and $WSe_2$) and hBN were prepared by mechanically exfoliating[22] their bulk crystals (natural $MoS_2$; CVT-grown $MoSe_2$, $WS_2$, and $WSe_2$ from 2D Semiconductors Inc.). Some $WSe_2$ samples were exfoliated from high-quality bulk crystals grown by a self-flux method.[23] Si substrates with 285 nm $SiO_2$ were treated with UV-generated ozone before the mechanical exfoliation. $WSe_2$ supported on multilayer hBN was fabricated by dry transfer method. Thermal oxidation was carried out in a mixed gas flow of $O_2$ (0.05 L/min) and Ar (0.20 L/min) using a quartz-tube furnace (Lindberg, Blue M) as described in a previous paper.[19] The temperature of the reactor was chosen in the range of 280 ~ 380 °C, and the reaction was continued for 2 h unless noted otherwise.

**Raman and photoluminescence spectroscopy.** The thickness and quality of each TMD sample were identified with differential reflectance[24] and Raman spectroscopy.[25] Raman and PL spectra were obtained with a home-built micro-Raman spectrometer setup described elsewhere.[26] Briefly, a laser beam operated at 457 nm was focused onto a sample with a spot size of ~1 μm using a microscope objective (40×, numerical aperture = 0.60). Backscattered signals were collected with the same objective and guided to a spectrometer equipped with a liquid nitrogen-cooled CCD detector. The overall spectral accuracy of Raman and PL measurement was better than 1 and 5 $cm^{-1}$, respectively. The average power on the sample surface was maintained below 0.2 mW to avoid photoinduced effects.

**AFM measurements.** The topological details of samples were investigated using an atomic force microscope (Park Systems, XE-70) before and after thermal oxidation. Height and phase images were obtained in a noncontact mode unless noted otherwise, whereas LFM images were obtained in a contact mode. Phase images gave much higher contrast for oxidized WX2 samples than height information. The nominal radius of Si tips was 8 nm (MicroMasch, NSC-15 for noncontact and CSC-17 for contact mode).

**SHG spectroscopy.** SHG measurements were performed with a home-built setup configured upon a commercial microscope (Nikon, Ti-U) as described elsewhere.[19] Briefly, the plane-polarized output from a Ti:sapphire laser (Coherent Inc., Chameleon) was used as a fundamental pulse with duration and repetition rate of 140 fs and 80 MHz, respectively. The beam was focused on samples with a spot size of ~ 2 μm using a microscope objective (40×, numerical aperture = 0.60). Backscattered SHG signals were collected with the same objective and guided to a spectrometer equipped with a thermoelectrically cooled CCD detector (Andor Inc., DU971P). An analyzing polarizer was placed in front of the spectrometer to select polarization components of interest. Samples were rotated with a precision of 0.2 degrees to vary the polarization angle of the incident fundamental beam.

**Results and discussion**

**Varying anisotropy in oxidation of four TMDs.** Single and few-layer TMD samples of four kinds ($MX_2$: M = Mo or W, X = S or Se) were prepared by mechanical exfoliation of commercial 2H-type bulk crystals (see Supplementary Note B for detailed methods). The thickness and quality of the samples were characterized using Raman spectroscopy, reflectance contrast, and atomic force microscopy (AFM) as described elsewhere.[19] Figures 1a ~ 1d show AFM images of four 1L TMDs obtained after oxidation at 280 ~ 380 ºC. Whereas their reactions may be represented by the following generic equation, $2MX_2 + 7O_2 \rightarrow 2MO_3 + 4XO_2$,[20, 21] the four TMDs varied drastically in their reaction-induced nanoscopic patterns. As shown in the height images of $MoS_2$ and $MoSe_2$ (Figs. 1a & 1b), oxidation led to the triangular etch pits (TEs) that were almost equilateral and aligned in the same direction. The TE areas in $MoX_2$ were $1.1 \pm 0.2$ nm lower than unreacted areas and revealed bare substrates as shown in their phase images (Fig. S1). The triangular shapes found in the phase images of $WS_2$ and $WSe_2$, however, were significantly rounded at their apexes (Figs. 1c & 1d). Remarkably, their height images revealed that the rounded triangles in $WX_2$ were protrusions slightly higher than unreacted regions (Figs. S1c & S1d). The substantial phase decrease[27] for the triangular areas indicated their chemical transformation. The optical bleaching and disappearance of Raman and photoluminescence peaks characteristic of $WS_2$ (Fig. S2) suggested that the triangles of $WX_2$ are tungsten oxides. Complete etching in $MoX_2$ indicates that most stoichiometric ($MoO_3$, $SO_2$, and $SeO_2$) and non-stoichiometric products are volatile at the reaction temperature, which is consistent with the fact that clusters of Mo oxides and suboxides sublime readily unlike bulk $MoO_3$.[28] Solidification as triangular oxides (TOs) observed for $WX_2$ is attributed to the higher thermal stability of W oxides (m.p. = 1473 °C) compared with Mo oxides (m.p. = 802 °C).[29] It is also to be noted that the density of TEs and TOs is much higher for $MSe_2$ than $MS_2$, which will be discussed later.

**SHG-determined crystallographic orientations of triangles.** Although less conspicuous because of their highly curved edges, TOs of $WX_2$ were also aligned in parallel. To determine the crystallographic orientation of the triangular patterns, we employed SHG spectroscopy, which detects the frequency-doubled signals of an incident fundamental beam. Because of their non-centrosymmetry and strong light-matter interaction, 1L of the four TMDs serve as efficient

SHG media.[30] For angle-resolved SHG measurements, samples were rotated by an azimuthal angle ($\theta$) with respect to the polarization direction of the fundamental beam ($\vec{E}_\omega$), while the SH field ($\vec{E}_{2\omega}$) parallel to $\vec{E}_\omega$ was selected (Fig. 1e). The symmetry of their second-order susceptibility tensor requires that the parallel SHG intensity ($I_{SHG}$) is proportional to $\cos^2 3\theta$ and reach maxima when $\vec{E}_\omega$ is parallel to the crystallographic armchair (AC) direction of the lattice.[30]

Figure 1f shows the polar graph of $I_{SHG}$ obtained from the oxidized WSe$_2$ sample. The data well fit with the predicted function indicated that one of the AC directions is at ~16º with respect to the horizontal axis of the polar graph. For the other TMDs, their AC directions were also determined with SHG measurements (Fig. S3) and designated with double-headed arrows in Fig. 1a ~ 1d. It can be readily seen that all the edges of TEs and TOs are parallel to zigzag (ZZ) directions that are normal to the AC lines. As depicted in Fig. 1e, two types of ZZ edges can be formed without considering lattice reconstruction: M-terminated (ZZ$_M$) and X-terminated (ZZ$_X$). As the controversy[20, 21] over the two possibilities was resolved by recent transmission electron microscopy studies,[31, 32] TEs and TOs found in this study were accordingly assigned to be ZZ$_M$-terminated.

**Top and bottom surfaces subject to reactions.** Unlike 1Ls, multilayered MoS$_2$ and MoSe$_2$ exhibited TOs instead of TEs (Figs. 2a & 2b). Whereas the TO in MoS$_2$ (Fig. 2a) spanned the 2L and 1L areas, the latter was completely etched away as shown in Fig. 1a. This indicates that TOs are formed by the oxidation of the bottommost layer contacting SiO$_2$ substrates. For TOs (Mo oxides including suboxides) to be formed in MoX$_2$ multilayers, oxygen molecules diffuse through the TMD-substrate interfaces and reach reaction centers at elevated temperatures.[19]

One type of product in the form of Mo oxides and their oligomers[28] does not diffuse efficiently because of their large size and remain within the triangle forming TOs. In contrast, chalcogen oxides ($SO_2$ or $SeO_2$) may diffuse out because of their small size. The TOs in $MoX_2$ are clearly visible in their phase images (Fig. S4), supporting that they are chemically distinct from the pristine areas. It is also notable that the vertices and edges of TEs are more rounded than those of TOs (Figs. 1a & 2a), which will be discussed below. Whereas TEs were also found in 2 ~ 4L $MoS_2$ and $MoSe_2$ as shown in Fig. S5, they were several times smaller than TOs. Unlike monolayer TEs (Fig. S1), the TEs of multilayer $MoX_2$ were not discernible in their phase images (Fig. S5), because etch-pit regions also contain and expose pristine $MoX_2$.

In multilayer $WS_2$ and $WSe_2$, TOs were clearly visible in the phase images (Fig. 2c & 2d) and found to be slightly elevated (Fig. S4c & S4d) as in 1L. As will be discussed below, however, their shapes are more rounded than those in 1L $WX_2$. Because W oxides formed in 1L $WX_2$ were found to survive the oxidation temperatures unlike Mo oxides, some of TOs may be located in the topmost layers of multilayer $WX_2$. To distinguish the exposed TOs from hidden TOs, we performed lateral force microscopy (LFM) for oxidized 2L $WSe_2$ showing multiple TOs in its height image (Fig. 2e & 2f). As depicted in Fig. 2g, the tip in a contact AFM mode essentially measures the friction against the surface of samples during LFM scans.[33] Whereas some TOs in 2L $WSe_2$ were conspicuous in their LFM image (Fig. 2f), the others could hardly be discerned from the unreacted areas unlike their height image. Because the surfaces of intact TMDs exhibit extremely low friction compared to oxides including silica substrates,[33] the low-friction (high-friction) triangles were assigned to $TO_B$ ($TO_T$) that were formed in the bottommost (topmost) layers. The effective frictional coefficient of $TO_T$ including those in 1L was four times higher than that of $TO_B$ (Fig. 2h). Notably, the average density was 10 times higher for $TO_B$ than $TO_T$ (Fig. 2e). This fact indicates that the reaction centers are more abundant in the bottommost layers, which will be justified below.

**Quantification of anisotropy and kinetic Wulff construction.** As depicted for 4L $MoS_2$ (Fig. 2a), we quantified the roundness of the triangular structures by the anisotropy index (R) defined as the ratio between the radii of the two circles that inscribes in and circumscribes on each triangle, respectively. R of 4L $MoS_2$ obtained from multiple TOs was $0.60 \pm 0.03$, which is noticeably higher than 0.5 for an equilateral triangle. Notably, the anisotropy index was higher for 1L than multilayers of $MoS_2$ (Fig. 2i). On the other hand, rounded TOs of 4L $WS_2$ resulted in a higher anisotropy value of $0.92 \pm 0.04$ and decreased for thinner layers as shown in Fig. 2i. Remarkably, $MoSe_2$ and $WSe_2$ exhibited thickness-dependent R values that are similar to $MoS_2$ and $WS_2$, respectively, which suggests that the anisotropic propagation of the reactions is mainly governed by the metallic elements of TMDs.

The material-dependent anisotropy observed in the oxidized 2D TMDs is dictated by the competition among different types of edges with varying oxidation rates. As depicted in Fig. 3a, the reaction centers of TEs or TOs at their early stage can be approximated as a tiny dodecagonal hole terminated with 6 armchair (AC) and 6 zigzag (ZZ) edges. One half ($ZZ_M$) of the latter exposes metallic atoms with the other half ($ZZ_X$) exposing chalcogens. We note, however, that the chemical structures given in Fig. 3a should be taken symbolically because multiple structures of similar formation energies are likely to exist for each of AC and ZZ edges.[31, 32, 34] Moreover, the edge structures of TEs and TOs can be affected by the oxidation that may induce reconstruction of edges.[20, 32] According to the kinetic Wulff construction (KWC) theory,[34-36] the edges with a higher rate (***k***) are consumed faster and become smaller in abundance. As the reaction proceeds further, the dodecagonal seed will grow and be terminated with the edges with the lowest $k$. Assuming that etch pits consist of the three types of edges, total 13 distinctive sets of ($k_{ZZM}$, $k_{ZZX}$, $k_{AC}$) that are different in the order of magnitude are possible as summarized in Fig. S6. The KWC theory predicts that each combination eventually leads to either of a triangle, hexagon, nonagon, or dodecagon as shown in Fig. 3b. It can be seen that the most anisotropic reactions generating triangles may result from multiple sets with either of $k_{ZZM}$ or $k_{ZZX}$ smallest. For the nondegenerate cases where the three rates are different from each other (cases 1 ~ 6 in Fig. S6), the starting dodecagon eventually turns into a triangle unless $k_{AC}$ is smallest. For the degenerate cases (cases 7 ~ 13 in Fig. S6), triangular patterns can also be formed when either type of ZZ edges is slowest in the reaction rate. Although the anisotropy index values given in Fig. 2i do not pinpoint which case each TMD belongs to, it can be concluded that $k_{ZZM}$ is smallest based on the aforementioned crystallographic assignment on edges of TEs and TOs. Moreover, the minimum/maximum ratio among the three rates is larger for $WX_2$ than $MoX_2$ because the anisotropy index of rounded triangles is determined by the ratio.

**Edge-resolved determination of reaction rates.** To corroborate the above model, we determined edge-specific oxidation rates of $MoSe_2$ and $WSe_2$. As shown in Fig. 4a and 4b, oxidation left residues along the original edges of the pristine TMDs. The assignment could be confirmed by the linear residues (marked by red arrows) that bisect the trench-like W oxide band (Fig. 4b). This fact also indicates that the 1L $WSe_2$ sample originally contained a crack line where the residues are located.[37] The oxidation rate of an edge was defined as the distance ($d$) between the two edges before and after the reaction divided by the duration of oxidation. For $MoSe_2$ and $WSe_2$, the degree of oxidation varied significantly depending on the direction of edges. In particular, $ZZ_{Se}$ edges that were anti-parallel to $ZZ_{Mo}$ of TEs in $MoSe_2$ underwent oxidative etching twice faster than $ZZ_{Mo}$ (Fig. 4a).

For quantitative understanding, angle-specific oxidation rate ($k_\theta$) was obtained for many samples as shown in Fig. 4c and 4d. To determine θ that was defined as the angle between the tangent to a given edge and AC directions (0 < θ < 120º) that are represented with the double-headed arrows within the AFM images. Between the two alternative ZZ directions, $ZZ_M$ was assigned to the one that was parallel to the edges of TEs or TOs as explained earlier. Fig. 4c showed that $k_\theta$ of $MoSe_2$ reached maximum and minimum at θ = 90 and 30º, representing $k_{ZZX}$ and $k_{ZZM}$, respectively, and intermediate values for 60º that corresponded to $k_{AC}$. Whereas $k_\theta$ of $WSe_2$ exhibited a similar pattern (Fig. 4d), its modulation among different edges was significantly reduced (maximum/minimum = 1.5).

These results that directly mapped anisotropic reactions agreed with observed triangular patterns of varying anisotropy index. Specifically, the triangles of $MoSe_2$ with small anisotropy index values can be well represented by the case No. 4 of Fig. S6 ($k_{ZZX}$ : $k_{AC}$ : $k_{ZZM}$ = 1 : 0.77 : 0.55). The increased isotropy in TOs of $WSe_2$ is consistent with the fact that the differences among the three $k$'s are less significant ($k_{ZZX}$ : $k_{AC}$ : $k_{ZZM}$ = 1 : 0.88 : 0.66) than those of $MoSe_2$. In addition, the experimental edge-specific rates allowed us to validate whether the current reactions are consistent with the KWC model. For further quantification, $k_\theta$ of $MoSe_2$ and $WSe_2$ were fitted with a sinusoidal function with a constant background (Fig. 4c & 4d). Then, each fit could be represented by its minimum/maximum ratio ($k_{ZZM}/k_{ZZX}$), denoted $R_E$. In Fig. 4e, we simulated oxidation patterns that would grow from a point reaction center for a given set of $k_\theta$ (red line) represented by $R_E$. In the KWC limit, the edges of TEs and TOs will be shaped by the inner envelope of the Wulff lines in black. For the larger $R_E$, KWC predicts the more rounded triangles. As shown in Fig. 4f, the anisotropy index (R) is essentially identical to $R_E$ when $R_E > 0.7$. As $R_E$ decreases to zero, the envelope approaches an equilateral triangle with the smallest R of 0.5. Notably, the experimental data in Fig. 4f suggest that $MoSe_2$ is consistent with the KWC prediction unlike $WSe_2$ which exhibited noticeable deviation.

**Chemical nature of reaction centers.** In Fig. 5a, metal selenides ($\sim 1 \times 10^9$ cm$^{-2}$) exhibited an order of magnitude denser TEs or TOs than metal sulfides ($\sim 1 \times 10^8$ cm$^{-2}$), which was valid irrespective of thickness. The robustness of sulfides is in line with their higher activation energy for dissociative adsorption of $O_2$ on chalcogen vacancies.[38] In early studies on $MoS_2$,[20, 21] it was inferred that oxidation started at preexisting defects and proceeded synchronously. However, the observed areal densities of TEs and TOs were orders of magnitude lower than those for major structural defects ($10^{12} \sim 10^{13}$ cm$^{-2}$) found in the TMDs.[13, 39, 40] In Fig. S7, we tested high-quality $WSe_2$ samples, grown in the self-flux method, that have 100 times fewer point defects than the conventional crystals grown by chemical vapor transport (CVT).[23] However, their TO density was not meaningfully different from that of the samples exfoliated from CVT-grown crystals. This indicates that the most abundant single-atom defects that were greatly reduced in the flux-grown $WSe_2$ are insufficient to serve alone as the reaction centers. As shown in Fig. 5b, no TOs were found in 2L $WSe_2$ samples supported on hBN substrates except at the hBN steps of 12 nm in height (Fig. S8). The density was estimated to be at least two orders of magnitude smaller on flat hBN than on $SiO_2$/Si substrates (Fig. S7). These results led us to conclude that the observed TEs and TOs are initiated at lattice sites that are deformed by the interaction with underlying substrates. This interpretation is consistent with the fact that the edges of $WSe_2$ underwent oxidation even when supported on hBN (Fig. S8c). We also note that TEs and TOs had fairly broad size distributions (Fig. S9), which result from unknown heterogeneity in the nucleation or propagation steps of the reaction. Mechanistic understanding of the initiation step requires further experimental and theoretical studies.

**Conclusion**

We reported material-dependent anisotropy in the thermal oxidation of four 2D TMDs. SHG spectroscopy combined with AFM method allowed us to quantify the reaction rates as a function of crystallographic direction. The reactions proceeded fastest (slowest) at chalcogen-terminated (metal-terminated) ZZ edges with AC edges in the middle. The degree of anisotropy was higher for W dichalcogenides than Mo counterparts and exhibited opposite dependence on thickness. We also show that the reactions are initiated at substrate-mediated defects that are located on the bottom and top surfaces of 2D TMDs.

## ASSOCIATED CONTENT

**Supporting Information.** Experimental methods, fitting function for edge-specific reaction rates; AFM height and phase images of 1L TMDs, Raman and photoluminescence spectra of 3L $WS_2$, determination of crystallographic orientation, complementary AFM images of four multilayer TMDs, TEs in multilayer $MoSe_2$, oxidation patterns predicted by the KWC theory, density of reaction centers, no reactions for $WSe_2$ supported on hBN, size distribution of TOs in 1L $WSe_2$; supplementary references.

## AUTHOR INFORMATION

**Corresponding Author**

*E-mail: sunryu@postech.ac.kr

**Author Contributions**

The manuscript was written through the contributions of all authors. All authors have given approval to the final version of the manuscript.

**Notes**

The authors declare no conflict of interest.

ACKNOWLEDGMENTS

This work was supported by the National Research Foundation of Korea (NRF-2020R1A2C2004865 and NRF- 2019R1A4A1027934).

**FIGURES & CAPTIONS**

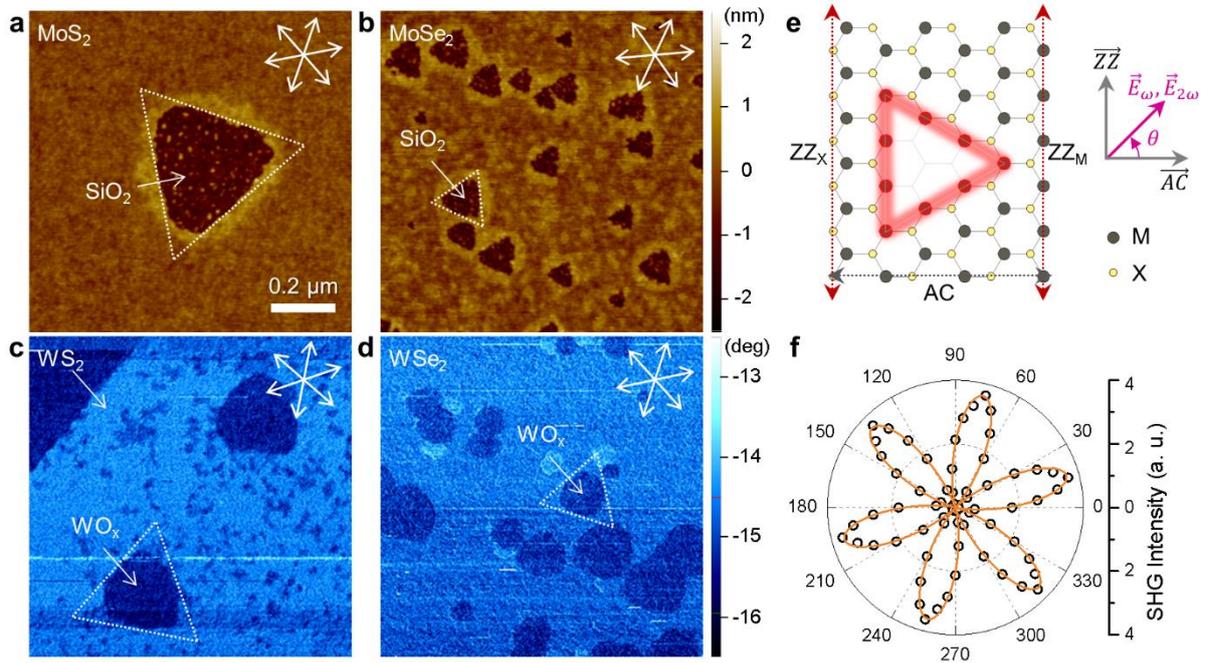

**Figure 1. In-plane anisotropy in thermal oxidation of 1L TMDs.** (a ~ b) AFM height images of triangular etch pits (TEs) of 1L MoS$_2$ (a) and MoSe$_2$ (b). (c ~ d) Phase images of triangular oxides (TOs) of 1L WS$_2$ (c) and WSe$_2$ (d). SHG-determined armchair directions are denoted by the double-headed arrows of 6-fold symmetry. Oxidation-generated triangular structures were shown with ZZ$_M$-edged triangles of white dotted lines. T$_{ox}$ of (a ~ d) was 330, 290, 360, and 300 °C, respectively. The size of each image is the same. (a ~ d) (e) Scheme for polarized SHG measurements of 1L TMDs with ZZ$_X$ (left edge), ZZ$_M$ (right edge), AC edges (top and bottom), and ZZ$_M$-terminated TE (highlighted with a red triangle). The angle ($\theta$) between AC edge and SHG signal ($\vec{E}_{2\omega}$), which was parallel to incident fundamental beam ($\vec{E}_\omega$), was varied for the determination of crystallographic orientation. (f) SHG intensity of oxidized 1L WSe$_2$ given as a function of $\theta$.

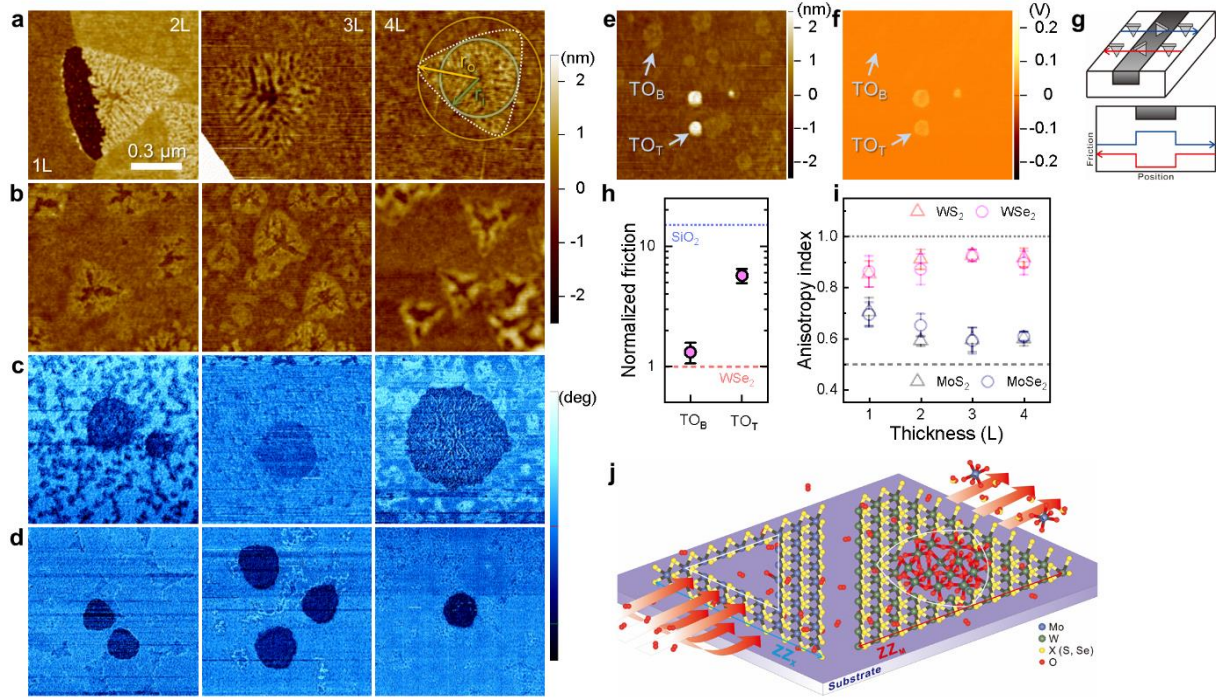

**Figure 2. Diverse reactions in multilayer TMDs.** (a ~ d) Height (a & b) and phase (c & d) images of oxidized $MoS_2$ (a), $MoSe_2$ (b), $WS_2$ (c), and $WSe_2$ (d): 2, 3, 4L in thickness from left to right. $T_{ox}$ and $t_{ox}$ are given in Fig. S4. The anisotropy index (R) of the TO in 4L $MoS_2$ (a) was defined by $r_i/r_o$, where $r_i$ ($r_o$) is the radius of the circle that inscribes in (circumscribes on) the TO. The vertical scale values of phase images (c & d) are from -7, -13, -16 (c), and -8, -18, -18 (d) to n - 4 degrees for 2, 3, 4L, respectively. (i.e., 2L WS2: -7° to -11°) (e & f) Height (e) and LFM (f) images of 2L $WSe_2$ ($T_{ox}$ = 300 °C and $t_{ox}$ = 2 h). The size of each image is the same. (a ~ f) (g) Scheme of LFM scans (top) and resulting frictional force profiles (bottom). (h) Frictional force of TOs in the top ($TO_T$) and bottommost ($TO_B$) layers of oxidized multilayer $WSe_2$. The red and blue lines represent pristine $WSe_2$ and $SiO_2$ substrates, respectively. (i) Anisotropy index of four TMDs as a function of thickness. Each data point was obtained by statistical treatment of multiple (up to 10) samples (typical area of each sample= ~50 μm²). (j) Reaction scheme for oxidation with differing anisotropy for Mo and W dichalcogenides: reactions may be initiated on the top and bottom surfaces of 1L and multilayers (1L is depicted for simplicity).

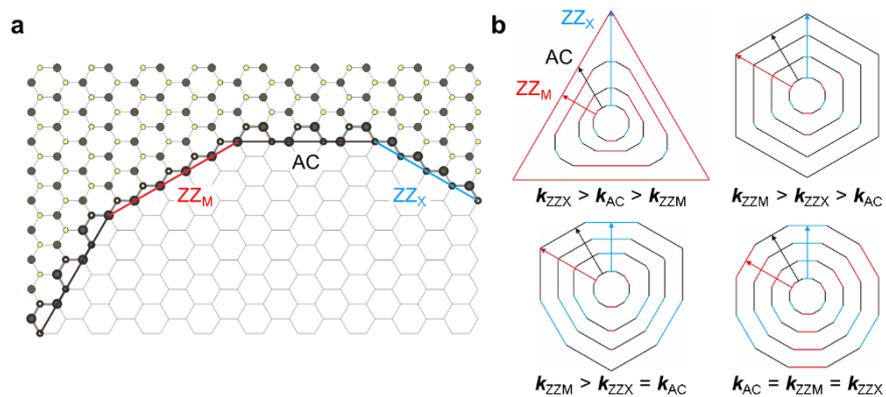

**Figure 3. KWC-based prediction of oxidative etching patterns.** (a) Schematic dodecagonal model as a reaction center with three representative edges: $ZZ_M$, AC, and $ZZ_X$ represented by red, black, and blue lines, respectively. (b) Kinetic Wulff construction for four sets of ($k_{ZZX}$, $k_{AC}$, $k_{ZZM}$) that lead to various polygonal etch patterns. The innermost dodecagons are the reaction centers with the length of each arrow representing the rate (*k*) of each edge. (see Fig. S6 for all possible combinations)

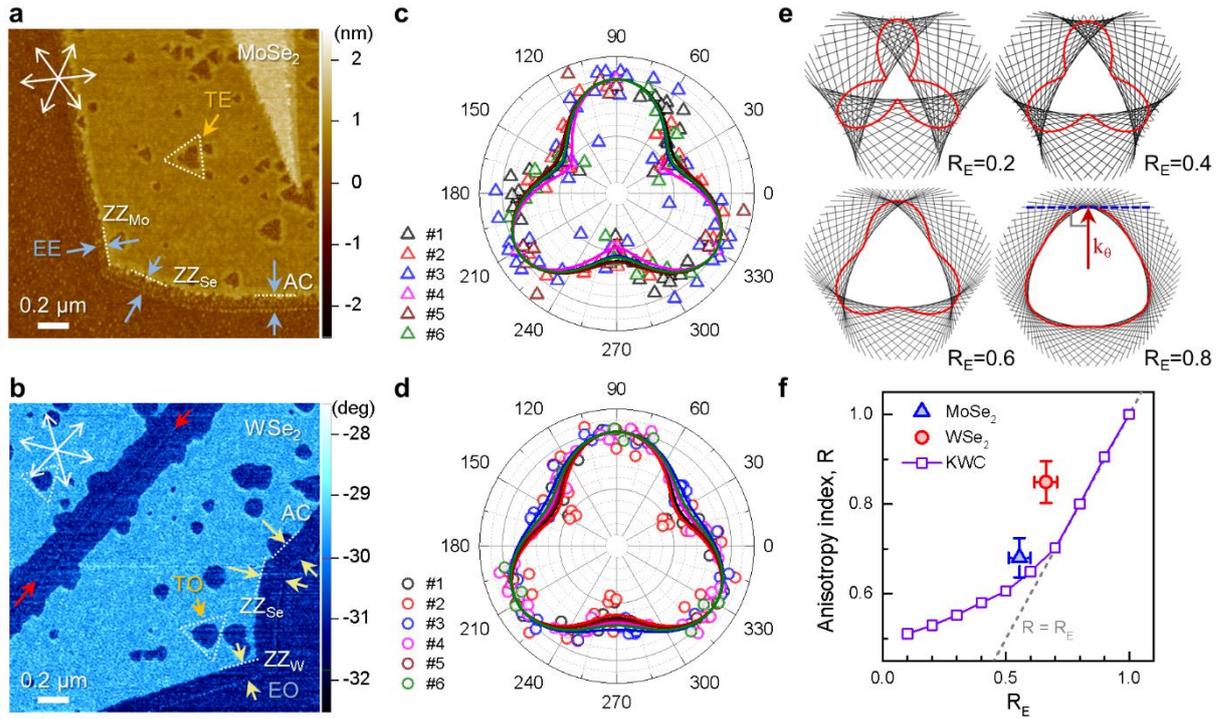

**Figure 4. Determination of edge-specific rates and kinetic Wulff construction.** (a ~ b) Height (a) and phase (b) images of oxidized 1L MoSe$_2$ (T$_{ox}$ = 290 °C, t$_{ox}$ = 2 h) and 1L WSe$_2$ (T$_{ox}$ = 300 °C, t$_{ox}$ = 2 h), respectively. In addition to TEs and TOs, the outer edges of MoSe$_2$ and WSe$_2$ underwent edge etching (EE) and edge oxidation (EO), respectively. The degree of edge reaction was marked by a pair of blue or yellow arrows. The differentiation between ZZ$_M$ and ZZ$_{Se}$ was made with respect to the orientation of TEs or TOs (see the main text). (c ~ d) Normalized reaction rate for EE of MoSe$_2$ (c) and EO of WSe$_2$ (d) obtained as a function of edge direction: 0, 30, and 90° respectively correspond to AC, ZZ$_M$, and ZZ$_X$ as defined in Fig. 1e. Because of the 3-fold symmetry of the lattice, data sets determined for 0 ~ 120° were repeated every 120° and fitted with a sinusoidal function (see Supplementary Note). Six samples (#1 ~ #6) were used for each TMD. (e) 2D KWC of oxidative etching simulated for the experimental edge-dependent reaction rates (red curve), where R$_E$ is the ratio of minimum to maximum rates for edge reactions. The kinetic Wulff shape is given as the interior envelope formed by the Wulff lines (in black), one of which in blue (R$_E$ = 0.8) is shown to be normal to $k_\theta$. (f) Relation between the anisotropy index R and R$_E$ obtained from (e). The experimental data for MoSe$_2$ and WSe$_2$ are given with the line for R = R$_E$.

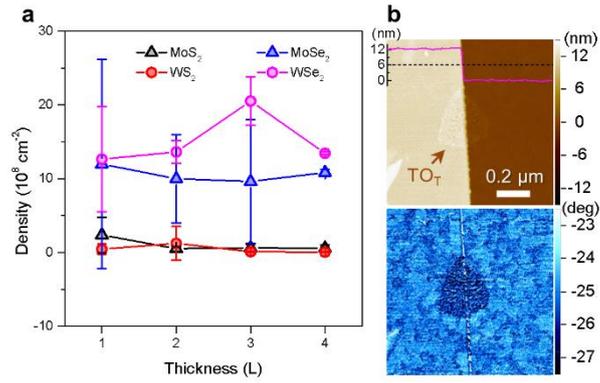

**Figure 5. Density and origin of reaction centers.** (a) Density of TEs and TOs for four TMDs of 1 ~ 4L. (b) Height (top) and phase (bottom) images of 2L WSe$_2$ supported on hBN. TOs were found only at the hBN step of ~12 nm in height as shown by the height profile in magenta. The size of each image is the same.